\documentclass[twocolumn,showpacs,preprintnumbers,amsmath,amssymb,pre]{revtex4-1}
\usepackage{color}

\usepackage{graphicx}
\usepackage{dcolumn}
\usepackage{bm}
\usepackage{subfigure}
\usepackage{amssymb}
\usepackage{multirow}
\graphicspath{{plots/}}

\begin{document}

\title{On ``Novel attractive forces'' between ions in quantum plasmas -- \\ failure of linearized quantum hydrodynamics}

\author{M. Bonitz, E. Pehlke, and T. Schoof}%
\affiliation{%
    Christian-Albrechts-Universit\"at zu Kiel, 
    Institut f\"ur Theoretische Physik und Astrophysik, 
    Leibnizstra\ss{}e 15, 24098 Kiel, Germany
}

\date{\today}

\begin{abstract}
In a recent letter [P.K. Shukla and B. Eliasson, Phys. Rev. Lett. {\bf 108}, 165007 (2012)] the discovery of a new 
attractive force between protons in a hydrogen plasma was reported that would be responsible for the formation of molecules and of a proton lattice. Here we show, based on ab initio density functional calculations and general considerations, that these predictions are wrong and caused by using linearized quantum hydrodynamics beyond the limits of its applicability.
\end{abstract}

\maketitle

\section{Introduction}\label{s:intro}
Recently Shukla and Eliasson published in Physical Review Letters \cite{shukla_11} a calculation where they applied Linearized Quantum Hydrodynamics in a simple way to obtain the effective potential between classical protons in a dense hydrogen plasma containing degenerate electrons. Their result indicated a novel attractive force between two protons. If well-founded, this is not only an interesting results, but the relatively simple calculation methodology is also potentially one of considerable importance as a theorist's tool. It is, therefore, important to discuss the implications of these results for real physical systems, such as dense quantum plasmas, and to rigorously verify the validity and limitations of the method against some more fundamental calculation. However, presenting such a result cannot be adequately done within the space limits of a Comment in Physical Review Letters, so it is being presented here as a full paper.

Shukla and Eliasson (SE hereafter) claim \cite{shukla_11} to have calculated their potential as that between classical protons in a dense hydrogen plasma containing degenerate electrons. The hydrogen plasma is considered in equilibrium (essentially in the ground state, $T=0$), and the authors stress that the observed ``novel attractive force'' is different from Friedel oscillations. Furthermore, from the existence of a ``hard core negative potential'', Eq.~(\ref{eq:phi_general}), they predict the formation of bound states, a proton lattice, as well as ``critical points and phase transitions at nanoscales''. If correct, these results could have far-reaching consequences for hydrogen and dense quantum plasmas, in general.

It is, therefore, of interest to consider the results of Shukla and Eliasson more in detail, which is the goal of this paper. We first summarize their approach -- Linearized Quantum Hydrodynamics (LQHD). Wen then re-evaluate their final formulas transforming to common length and energy scales. We then compare their screened proton potential to results from {\em ab initio} density functional (DFT) simulations and observe qualitative deviations. From this, we have to conclude that the LQHD results of Ref. \cite{shukla_11} are not applicable to dense quantum plasmas ad do not provide evidence for attractive interactions between protons. We conclude by analyzing the origin of the deviations of the results of Ref. \cite{shukla_11} from DFT which are traced to violation of the applicability range of linearized quantum hydrodynamics.

Dense correlated quantum plasmas are presently of high interest in many fields including condensed matter, astrophysics and laser plasmas, e.g. \cite{kremp-book,bonitz_pop08}. 
Despite remarkable theoretical and experimental progress over recent decades, even for the simplest plasma system -- hydrogen -- still interesting questions remain unsolved, including details of the phase diagram and the behavior under high compression, e.g.~\cite{silvera10,morales10,ashcroft12,galli04}. The modification of the pair interactions between the ions 
by the surrounding plasma is of prime importance for the theoretical understanding of these systems. In contrast to the familiar Coulomb potential, $\phi^i(r)=Q/r$, of an ion observed in vacuo,
in a plasma, correlation and quantum effects cause screening. At weak non-ideality this gives rise to an isotropic Yukawa-type potential, 
$\phi^i_s(r)=\frac{Q}{r} e^{-r/l_s}$, where $l_s$ is the screening length given, in the limit of a classical high-temperature plasma, by the Debye radius or, in a high-density quantum plasma, by the Thomas-Fermi length $L_{\rm TF}$. More general screened potentials are successfully computed using linear response theory to obtain the dynamic dielectric function, $\epsilon^l({\bf k},\omega)$ (longitudinal density response), giving rise to \cite{diel_function}
\begin{equation}
 \phi^i_s({\bf r}) = \frac{Q}{2\pi^2} \int d^3k \, \frac{e^{i{\bf k \cdot r}}}{k^2 \epsilon^l({\bf k}, 0)},
 \label{eq:phi_general}
\end{equation}
where the longitudinal dielectric function derives from the dielectric tensor via $\epsilon^l({\bf k},\omega) = \sum_{ij} k_i k_j\epsilon_{ij}({\bf k},\omega)/k^2$. The potential (\ref{eq:phi_general}) typically decays slower than the exponential Yukawa potential. 

Moreover, screening effects are known to give rise to an attractive 
region of the potential (``anti-screening'') in nonequilibrium situations such as a charge embedded into a streaming flow of oppositely charged particles, which is well studied theoretically \cite{shukla_rmp,hutchinson11,patrick_njp12}, and the resulting attractive force has been confirmed e.g. in dusty plasma experiments with streaming ions. 
No attractive forces have been seen in simulations in equilibrium plasmas with two exceptions: i) The formation of bound states (such as hydrogen molecules or molecular ions) evidently corresponds to a net attractive force between the constituents (hydrogen atoms, protons). ii) Oscillatory potentials with (very shallow) negative parts (Friedel oscillations) emerge in a strongly degenerate Fermi gas as a consequence of the Fermi edge singularity (essentially due to the step character of the zero temperature momentum distribution of an ideal Fermi gas), e.g.\ \cite{kittel}. Friedel oscillations have been observed in experiments probing surface states at very low temperature, e.g. \cite{friedel_exp}.

\section{Linearized quantum hydrodynamics (LQHD)}\label{s:lqhd}
In Ref.~\cite{shukla_11} SE compute the potential (\ref{eq:phi_general}) of a proton in a hydrogen plasma by 
\begin{enumerate}
 \item using zero temperature classical hydrodynamic equations for the electron density and mean velocity, coupled to the Poisson equation for the electrostatic potential,
 \item adding, in the momentum equation, quantum diffraction effects via the Bohm potential $V_B$ (quantum hydrodynamics, QHD),
 \item adding the pressure of the ideal Fermi gas at zero temperature,  
 \item adding exchange and correlation effects via an additional potential $V_{xc}$ using a simple parametrization, Ref. 33 in \cite{shukla_11},
 \item neglecting dynamic effects, $\epsilon^l({\bf k},\omega) \rightarrow \epsilon^l({\bf k},0)$,
 \item solving the resulting hydrodynamic equations in linear response for the dielectric function $\epsilon^l({\bf k},0;r_s)$ ($D$, in their notation), 
\end{enumerate}
which parametrically depends on the coupling parameter (Brueckner parameter) $r_s={\bar r}/a_B$ that completely defines the plasma state at zero temperature. [Here we introduced the mean interparticle distance, ${\bar r}$, which is related to the unperturbed density by $4\pi {\bar r}^3/3 = n^{-1}$].
For the explicit form of the used LQHD equations we refer to Ref.~\cite{shukla_11}, for the discussion of $V_{xc}$ and earlier related work, see Ref.~\cite{manfredi08}. In the following we restrict ourselves to the linearized equations used by SE, i.e. to the LQHD. The full QHD is beyond the scope of this paper, for references, see Ref.~\cite{shukla_11}.

\begin{figure}[h]
 \includegraphics[scale=0.85]{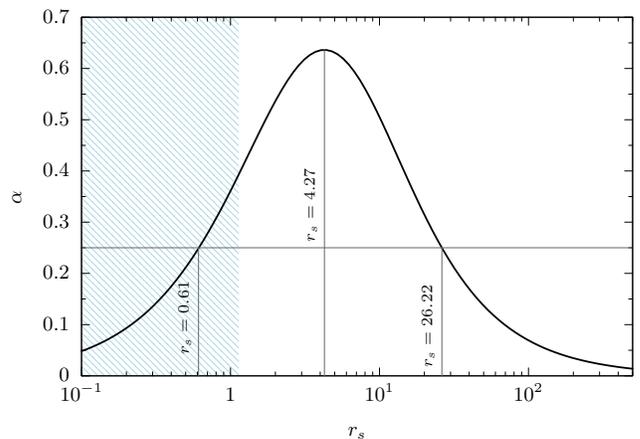}
\caption{
Coupling parameter $\alpha$ versus Brueckner parameter $r_s$. 
An attractive proton potential is predicted in the density interval 
$26.22 \ge r_s \ge 0.61$ at zero temperature. 
The shaded area denotes the range of $r_s$ where the plasmon energy is smaller than the Fermi energy, 
$\hbar \omega_{\rm pe} < k_{\rm B} T_{\rm F}$ (weak coupling).}
\label{fig:alpha}
\end{figure}

\begin{figure}[t]
 \includegraphics[scale=0.85]{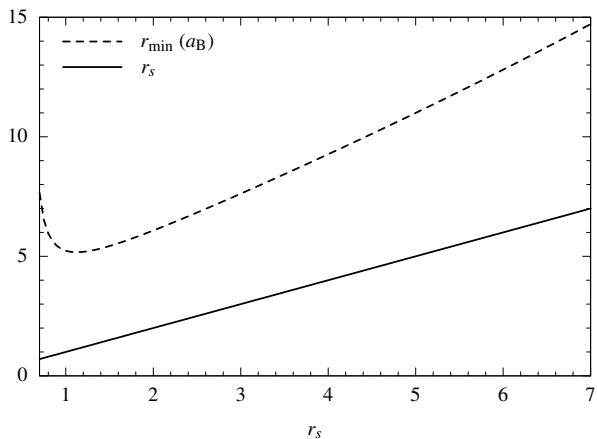}
\caption{Position of the attractive potential minimum of SE, Ref.~\cite{shukla_11}, in units of the Bohr radius as a function of $r_s$, compared to the mean interparticle distance. Note that the ground state nearest neighbor distance of two protons in the molecular phase ($r_s \gtrsim 2\dots 3$) is in the range of $1.5 \dots 2 a_B$.
} 
\label{fig:r_min}
\end{figure}

\begin{figure}[t]
 \includegraphics[scale=0.85]{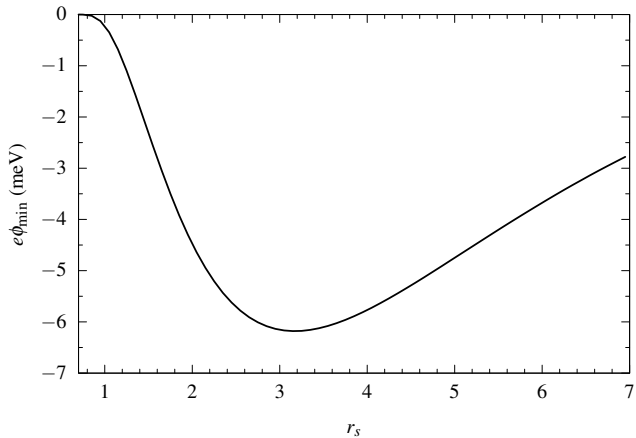}
\caption{Depth of the attractive potential minimum of SE, Ref.~\cite{shukla_11}. 
The deepest value is observed for $r_s\approx 3$ corresponding to $\alpha\approx 0.6$, cf. Fig.~\ref{fig:alpha}. 
} 
\label{fig:phi_min}
\end{figure}

Using LQHD, SE evaluate the static dielectric function, $\epsilon^l({\bf k},0)$, and screened proton potential,  $\phi^i_s$, at zero temperature, as a function of a single parameter $\alpha$. 
We plot $\alpha$ as a function of the Brueckner parameter in Fig.~\ref{fig:alpha}. SE observe that the screened potential of a proton, Eq.~(\ref{eq:phi_general}), develops a negative minimum for $\alpha > 0.25$. As can be seen in the figure, 
$\alpha$ exceeds the value $0.25$, in a range of coupling parameters $0.61 \le r_s \le 26.22$. 
\begin{figure}[h]
 \includegraphics[scale=0.85]{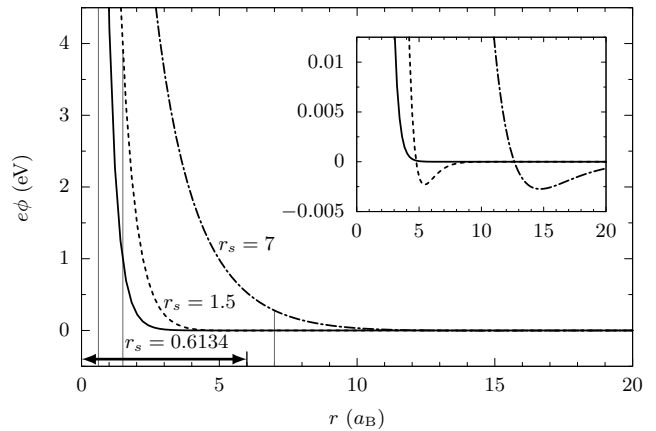}
\caption{Screened proton potential of SE for three densities. 
 A single negative minimum is observed, shown more clearly in the inset. The location of the minimum and its depth are shown 
 in Figs.~\ref{fig:r_min} and \ref{fig:phi_min}, respectively. Thin vertical lines indicate the equilibrium nearest neighbor distance of 
 two protons, cf. Fig.~\ref{fig:r_min}. The black arrow marks the range of values of $r$ shown in Fig.~\ref{fig:DFT-electrostatic_potential}.} 
\label{fig:pot_se}
\end{figure}
Let us now analyze the results for the screened proton potential in the whole range of $r_s$ values where an attractive hydrogen interaction is observed, as done in Ref.~\cite{shukla_11}.
In Fig.~\ref{fig:pot_se} we plot the potential energy corresponding to the screened proton potential in units of eV for three densities corresponding 
to $r_s=7, 1.5, 0.61$. At low and high densities (i.e. for cases where $\alpha < 0.25$), we confirm that the potential is purely repulsive. In agreement with Fig.~\ref{fig:alpha}, for $\alpha>0.25$ a 
negative minimum develops at a distance of several Bohr radii from the proton. The position of the minimum is plotted versus $r_s$ in Fig.~\ref{fig:r_min}. 
There exists a minimum of this distance of about $5a_B$ for $r_s\approx 1$ with a slow (rapid) increase for lower (higher) densities. Finally, we 
analyze the depth of the attractive potential minimum. 
These values are plotted in Fig.~\ref{fig:phi_min} for $r_s$ between $7$ and $0.6$ below which the minimum vanishes. The deepest minimum is observed around $r_s=3$ and amounts to about $6$meV corresponding to a temperature of about $65$K.
\section{Comparison of LQHD to ab initio DFT results}\label{s:dft}
We now compare the LQHD results of SE to what is known about dense hydrogen. 
There have been numerous studies of hydrogen plasmas based on analytical approximations and first-principle simulations as well, e.g. \cite{kremp-book,morales10,galli04}. There is hardly any question that the low-density ground state consists of $H_2$-molecules (either in a gas, liquid or solid phase). 
When the density is increased, the molecular binding is reduced until, eventually around $r_s \sim 2\dots 3$, molecules break up. 
Even though the detailed scenario maybe quite complex, e.g. \cite{morales10,ashcroft12}, involve intermediate states 
(including $H$-atoms) and vary with temperature
it is clear that, around some critical density $n_M$ (Mott density), all bound states will break up into electrons and protons, 
e.g. \cite{bonitz_prl05}. This pressure ionization is a gradual process and a reasonable estimate for $n_M$ corresponds to 
$r_s \approx 1.2 \dots 1.5$. Upon further density increase one expects formation of a liquid and, eventually, solid phase of protons 
embedded into a Fermi gas of electrons, e.g. \cite{bonitz_prl05,militzer06}. While there remain interesting questions about the precise 
values of the different phase boundaries \cite{filinov12}, the general structure of the hydrogen phase diagram at high pressures is well 
understood, e.g. \cite{silvera10}.

As an illustration, and for direct comparison with SE, we present results for the screened proton potential and for the 
interaction of hydrogen atoms embedded in a jellium background, as obtained from density-functional total energy calculations.  
The {\it ab initio} scheme of DFT rests on firm theoretical grounds and has been thoroughly tested in recent decades, in particular 
in application to dense plasmas, see e.g. refs.~\cite{morales10,militzer12}. 
It provides a fully consistent treatment of quantum effects 
and uses the fundamental Coulomb interaction between charged particles as input. 
The only uncontrolled approximation contained in DFT total-energy calculations is the approximation 
applied to the exchange-correlation energy functional, but the accuracy and limitations of the different approximations 
available have been investigated in detail for many applications in solid state physics and molecular chemistry~\cite{dft-overview}.
Therefore, DFT total-energy calculations can serve as a benchmark for the linearized quantum hydrodynamic model of SE. 
Before proceeding with our own calculations we point out that detailed DFT calculations have been 
carried through for H and H$_2$ in jellium before, see, e.~g., the work by Almbladh {\it et al.}, Ref.\ \cite{almbladh76}, 
by Bonev and Ashcroft, Ref.\ \cite{ashcroft01}, and by Song, Ref.\ \cite{song}. Our results concur with their findings. For other recent DFT results for dense hydrogen, see Refs.~\cite{militzer12,redmer,galli04} and references therein. 

In the following we restrict ourselves to spin-unpolarized simulations. It is a well-known artefact of
the usually applied approximations to the exchange-correlation energy functional that in vacuo the hydrogen 
molecule undergoes a transition from a spin-unpolarized to a spin-polarized ground state when the separation of the 
two protons is increased \cite{perdew95,fuchs05}. However, in their spin-polarized 
DFT-calculations for hydrogen atoms immersed in jellium, Nazarov {\it et al.} \cite{nazarov05} have found 
a spin-unpolarized ground-state for $ 3 < r_s < 14$. For this reason, and as no spin-polarization is to be 
expected at even higher electron densities, we have carried through only spin-unpolarized relaxations for 
$r_s = 1.5$ and $r_s = 7$. This is consistent with Ref.\ \cite{ashcroft01}.
\begin{figure}[t]
 \includegraphics[width=85mm]{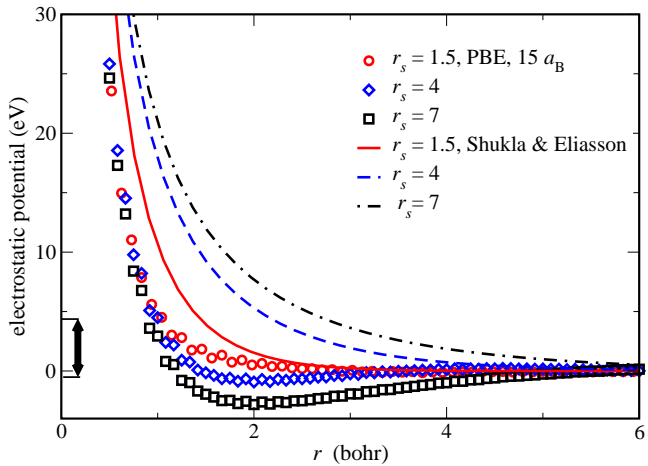}
\caption{(color online)
Electrostatic potential around an H-atom immersed in jellium. $r$ is the distance from the proton. DFT data have been 
calculated in a cubic box (size $ 15 a_0$) using the PBE-GGA for the exchange-correlation energy-functional, a single k-point, 
and a plane-wave cutoff-energy of 200 Ry ($r_s = 1.5$) or 300 Ry ($r_s = 4$ and $r_s=7$). The DFT data (symbols) are 
compared to the electrostatic potential from LQHD of Shukla and Eliasson, Ref.~\cite{shukla_11} (lines). The black arrow marks 
the voltage range shown in Fig.~\ref{fig:pot_se}.  
} 
\label{fig:DFT-electrostatic_potential}
\end{figure}
\begin{figure}[t]
 \includegraphics[width=85mm]{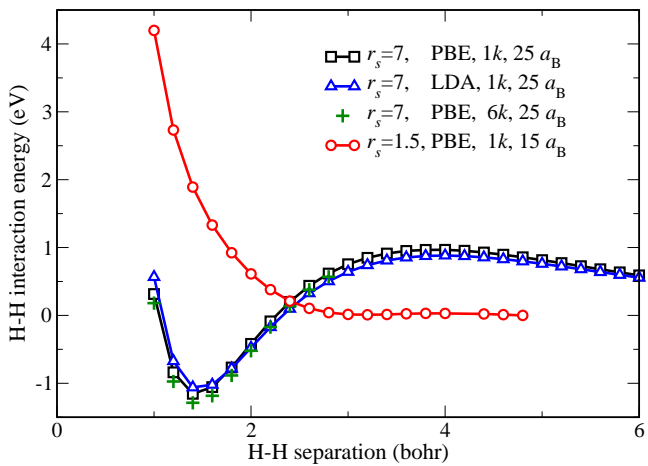}
\caption{ (color online)
Interaction energy of two H-atoms immersed in jellium for two densities. While for $r_s=7$ a minimum around the hydrogen molecule bond length in vacuo ($1.4$ bohr) is observed, for $r_s=1.5$, the molecular bound state is unstable. The DFT data have been calculated in a cubic box 
(with a size as noted in the inset) using PBE-GGA or LDA for the exchange-correlation energy-functional, a single
or 6 k-points in the irreducible part of the Brillouin zone, and a plane-wave cutoff-energy of 100 Ry. 
} 
\label{fig:DFT-interaction_energy}
\end{figure}

The data in Figs.\ \ref{fig:DFT-electrostatic_potential} and \ref{fig:DFT-interaction_energy} have been 
computed with the {\tt fhimd} code \cite{bockstedte97}. A $1/r$ Coulomb potential (i.e.\ no pseudopotential)  
has been taken for the proton--electron interaction. The cut-off energy of the plane-wave basis-set has been 
chosen at 100 Ry for the calculation of the H--H interaction energy, while larger values have been used for 
the calculation of the screened potential of an H-atom, or proton, in jellium. 
The two hydrogen atoms forming the H$_2$ molecule are put into a cubic super-cell with the length of the edge equal 
to 25 bohr (15 bohr), in case of $r_s = 7$ ($r_s = 1.5$). The super-cell is repeated periodically 
in all three directions. Calculations have been carried through using only a single special k-point. 
Additional computations with 6 special k-points derived from a $4 \times 4\times 4$ 
back-folding \cite{bockstedte97,monkhorst76} allow an error estimate, as can be seen in 
Fig.\ \ref{fig:DFT-interaction_energy}. The smearing of the Fermi distribution is 
small, $k_{\rm B}T = 0.025$ eV  or 0.1 eV, and the total energy has been extrapolated to zero temperature 
under the usual assumption of a linear heat capacity \cite{bockstedte97}. The generalized gradient 
approximation by Perdew and Ernzerhof (PBE96) \cite{PBE96} and, for comparison, the local-density 
approximation, have been applied to the exchange-correlation energy functional.   

Let us now consider the results. As a typical low-density case, we show in Fig.~\ref{fig:DFT-electrostatic_potential} the electrostatic 
potential around a proton in jellium for $r_s=7$. It exhibits a clear negative minimum around a distance 
of $r_{min} \approx 2 a_B$. The minimum becomes shallower with increasing electron density, cf. curve for $r_s=4$, and vanishes 
for $r_s=1.5$. There are further Friedel oscillations at larger distances, which, however, are often not well resolved due to computational limitations and the small size of the super-cell, but they are of no relevance for hydrogen at these conditions, see below. 
Our data can now be directly compared to the LQHD results of SE, shown by lines in Fig.~\ref{fig:DFT-electrostatic_potential}. 

In transferring the Shukla-Eliasson results from Fig.~\ref{fig:pot_se} to Fig.~\ref{fig:DFT-electrostatic_potential} one should note the large changes in scale and range. The horizontal axis in Fig.~\ref{fig:DFT-electrostatic_potential} extends out only to $6a_B$, while that of Fig.~\ref{fig:pot_se} extends more than three times further. Also the vertical scale of the potential energy in Fig.~\ref{fig:pot_se} is roughly six times smaller than that of Fig.~\ref{fig:DFT-electrostatic_potential}. Thus, on the scale of Fig.~\ref{fig:DFT-electrostatic_potential}, the LQHD minima, which only appear clearly in the inset of Fig.~\ref{fig:pot_se}, involve such small potential changes on the scale of Fig.~\ref{fig:DFT-electrostatic_potential} that they cannot be seen. The small dip in the LQHD potential for $r_s=1.5$ is at about $5.4 a_B$ near the right hand edge of the $r$-axis, while the dip for $r_s=7$ is even further away, far beyond the right hand edge. In any case, in the region of the DFT minimum, the LQHD potential is monotonically decreasing, i.e. without minima, and far from that of the full calculation, as denoted by the symbols.

In part we ascribe the minimum of the electrostatic potential from the DFT calculation in Fig.~\ref{fig:DFT-electrostatic_potential} to the formation of an H$^-$ ion, i.~e.\ 
a bound state which Almbladh {\it et al.} \cite{almbladh76} have found for $r_s > 1.9$. 
As pointed out by Bonev and Ashcroft \cite{ashcroft01}, this formation of a negative ion 
sheds doubts on the 
adequateness of jellium-type (or one-component plasma) models  for the description of a nonideal hydrogen plasma.
The recent ab initio simulations of hydrogen by Morales et al. \cite{morales10} and Bonev et al. \cite{galli04} are in fact based on DFT-Born-Oppenheimer molecular dynamics simulations of ensembles or Monte Carlo simulations and thus fully account for the dynamics of the protons.

The interaction energy of two hydrogen atoms immersed in a jellium background with $r_s=7$ or $r_s=1.5$ 
is displayed in Fig.\ \ref{fig:DFT-interaction_energy}. In fact, the H--H interaction energy has been computed
by Bonev and Ashcroft \cite{ashcroft01} before, using the DFT total-energy program {\tt VASP}. They 
point out that linear response theory would be inadequate for this purpose. Furthermore, they  
describe the fate of the hydrogen bond when the electron density is increased: For $r_s > 3$ 
 a local energy minimum at a H--H separation larger, but 
still close to, the bond length of the hydrogen molecule in vacuum, and an energy barrier 
towards H$_2$ association are observed \cite{ashcroft01}. Our DFT results summarized in Fig.\ \ref{fig:DFT-interaction_energy} 
show the same behavior: For $r_s = 7$ we obtain a stable H$_2$-bond at a bond length comparable 
to the bond length of a hydrogen molecule in vacuo, but with a much smaller binding energy. 
This result is not sensitive with respect to the approximation applied to the exchange-correlation 
functional. Also the more technical effect of the restricted k-point sampling can be read 
from the Figure. Comparing again to the LQHD results of Shukla and Eliasson, it is obvious
that the interaction energy minimum found in the {\it ab initio} data is not included 
in their approach. Thus, the SE screened proton potential has, 
for small densities (e.g.\ $r_s=7$), a qualitatively incorrect shape, in particular it 
completely misses the molecular bound state. 

For larger density, i.~e.\ for $r_s=1.5$, Fig.~\ref{fig:DFT-interaction_energy} shows that the minimum 
in our simulations becomes very shallow or vanishes at all, within the accuracy of the numerical DFT computations.     
Similarly, we do not resolve extremely shallow (below $100$meV) 
binding potentials at larger distances from the proton, such as the ones related to Friedel oscillations. 
These oscillations could be reproduced by DFT using a substantial computational effort. 
However, there is no need for this since such extremely small features of the potential 
are expected to be of no relevance for all dense plasma applications since they always encounter a finite electron 
temperature of at least one eV. 

\section{Discussion. Failure of LQHD}\label{s:dis}
%
The approach by SE in Ref.~\cite{shukla_11} starts from the assumption that the ions are immobile and embedded into a neutralizing background (see Eq.~(3) in Ref.~\cite{shukla_11}). We have thus used DFT, together with the LDA or the PBE-GGA \cite{perdew95} applied to the exchange-correlation energy functional, to calculate the interaction energy of two protons immersed in a jellium background, without any further approximation applied to the quantum-mechanical problem. For the reliability of the DFT as compared to Quantum Monte Carlo simulations we refer e.g. to Ref.~\cite{militzer12}. Thus the DFT results can serve as a reference to evaluate the accuracy of the interaction potential as derived from the LQHD approach by SE \cite{jellium_comment}. 

As shown from the comparison to ab initio DFT simulations, the SE potential 
completely misses the bound states of protons in low temperature hydrogen.
When the density is increased the deviations between LQHD and DFT potentials become smaller but are still noticeable.

Yet even at higher densities where no molecules exist the SE potential is qualitatively wrong because it does not show an attractive minimum for  $r_s < 0.61$ at all. Dielectric theories of the electron gas and electron liquid including correlations in the frame of local field corrections, however, confirm that Friedel oscillations persist up to high densities, e.g. \cite{simion,mahan}.
The present LQHD model of SE also misses the Friedel oscillations, as the authors themselves underline. 

Furthermore, we consider two limiting cases of the LQHD potential discussed by SE \cite{shukla_11}: first, in the limit $\alpha \to 0$ SE recover a Yukawa potential with the screening length $L_{\rm TF}$. This is correct for high densities,  $r_s \to 0$, but not for low densities, $r_s \gg 1$ where $\alpha \to 0$ as well, cf. Fig.~\ref{fig:alpha}. Second, SE recover, for $\alpha \gg 1$, the exponential cosine-screened Coulomb potential \cite{shukla_11}. This limit is questionable since $\alpha$ cannot exceed $0.65$, cf. Fig.~\ref{fig:alpha}.
These failures rule out any reliable predictions such as novel potential minima. Besides, the extremely 
low value of the associated binding energy (cf. Fig.~\ref{fig:phi_min}) would require a particularly accurate theory and careful verification.

Finally, let us analyze possible reasons for the failure of the linearized QHD model in application to dense two-component plasmas.
\begin{description}
 \item[i.] As any fluid theory, LQHD cannot resolve distances below some cut-off $r^*$. This has been pointed out by Manfredi and co-workers \cite{manfredi08} who performed a test of a 1D version of nonlinear QHD for thin metal films \cite{manfredi08}.
 \item[ii.] Shukla and Eliasson note \cite{shukla_11} that QHD is valid only for $\hbar \omega_{\rm pe} < k_BT_{\rm F}$, i.e. for 
weak coupling. This range is indicated by the shaded area in Fig.~\ref{fig:alpha}. Since the present model of LQHD contains 
exchange and correlation contributions, this range may actually be larger \cite{manfredi08}.
 \item[iii.] Linearized QHD fails when the linearization conditions are violated, i.e. when the perturbed density is comparable with the unperturbed density which is common in the description of strong attractive interactions.
\end{description}
While item (ii) is presently open and requires further analysis, our DFT simulation results allow us to directly verify (i.).
In a hydrodynamic description of classical plasmas the smallest length scale that can be resolved is the Debye length. Similarly, in the quantum case, this cut-off is expected to be of the order of the Thomas-Fermi length, 
$L_{\rm TF}=\hbar k_{\rm F}/(\sqrt{3} m_e\omega_{\rm pe})=(\pi/12)^{1/3}r_s^{1/2}a_B$. In fact, re-examining the DFT and LQHD data for the screened proton potential in Fig.~\ref{fig:DFT-electrostatic_potential} we observe that the most dramatic deviations occur on length scales smaller than $r^{*} \approx (3\dots 4) L_{\rm TF}$. Thus we confirm that LQHD cannot, by construction, yield potentials with atomic-scale resolution in a dense quantum plasma \cite{larger_distances}.

Yet the most severe limitation is, apparantly, related to the linearization of the QHD equations, item (iii).
In fact, similar linear response calculations for a proton in a degenerate electron gas have been done 
long ago. Almbladh et al. \cite{almbladh76} have performed density functional calculations and compared 
them to a linearized version of DFT (which, by construction, is essentially more accurate than any 
hydrodynamic approach). They observed a complete failure of the linear theory in the prediction of the 
electron density screening the proton. Not only is the electron density at the proton a 
factor $2$ to $33$ (for $r_s=1$ and $r_s=6$, respectively) too small, but linear theory generally 
predicted Friedel oscillations to occur at much too high distances from the proton. Thus, a possible explanation for the potential minimum observed in the present linearized QHD, cf. Fig.~\ref{fig:pot_se}, is that it is a trace of the Friedel oscillations (the first minimum) that is displaced to higher distances from the proton due to the linearization, cf. also Fig.~\ref{fig:r_min}. It would, therefore, be interesting to test this hypothesis by comparing the LQHD results of SE to full nonlinear QHD.

Let us assume for a moment that the attractive potential of SE would be a real effect and consider its implications for proton crystallization \cite{shukla_11}. To this end, we plot in Fig.~\ref{fig:r_min} the location of the potential minimum and compare it to the mean interparticle distance for a given density. It is obvious that there is a striking mismatch. Even if 
the system would be at zero temperature protons could not occupy the minimum locations simply because their density is several orders of magnitude too high (see also Fig.~\ref{fig:pot_se} where the mean interparticle distance is marked on the SE potential), making such a state energetically impossible. We thus have to conclude, based on DFT simulations and general considerations about the location and low depth of the potential minimum, that the predictions of SE of novel attractive forces, novel ion lattices, atoms, molecules, critical points and phase transitions in dense hydrogen are invalid.


\section*{Acknowledgements}
This work was supported by the Deutsche Forschungsgemeinschaft via SFB-TR 24 project A5. We are grateful fo the referees for suggestions that have helped improve the clarity of the manuscript.


\begin{thebibliography}{33}
\bibitem{jellium_comment} This, however, is not meant to imply that we suggest the jellium model to be a viable model to describe interactions in dense plasmas.
\bibitem{larger_distances} For completeness we mention that for distances larger than $r^*$ there are still noticeable deviations 
of the LQHD potential from the DFT result that are significantly larger than the minimum of the LQHD potential, cf. Fig.~\ref{fig:phi_min}.
\end{thebibliography}
\end{document}